\documentclass[journal,onecolumn]{IEEEtran}
\usepackage{times}
\usepackage{cite}
\usepackage{color}
\usepackage{epsf}
\usepackage{epsfig}
\usepackage{graphicx}
\usepackage{graphics}
\usepackage{epstopdf}
\usepackage[footnotesize]{caption}
\usepackage{amsmath}
\usepackage{amssymb}
\usepackage{amsthm}
\usepackage{amsxtra}
\usepackage{subcaption}
\usepackage[ruled]{algorithm2e}
\usepackage{algorithmic}
\usepackage{enumerate}
\usepackage{multirow}
\usepackage{enumerate}
\usepackage{amssymb}
\usepackage{lipsum}
\usepackage{setspace}
\usepackage{multirow}
\usepackage[normalem]{ulem}
\usepackage{color,soul}
\usepackage[table]{xcolor}
\usepackage{bm}

\newtheorem{definition}{\bf Definition}
\newtheorem{remark}{Remark}

\usepackage{rotating} 
\usepackage{graphicx} 

\newcommand{\Rmnum}[1]{\expandafter\@slowromancap\romannumeral #1@}

\newcommand{\ds}[1]{{\setlength{\parindent}{0mm}\textcolor{red}{#1}}}

\makeatother
\begin{document}
\title{Policy Design for Controlling Set-Point Temperature of ACs in Shared Spaces of Buildings}
\author{Wayes~Tushar$^*$,~Wang~Tao,~Lan~Lan,~Yunjian~Xu,~Chathura Withanage,~Chau~Yuen,~and~Kristin L. Wood
\thanks{$^*$Corresponding author: wayes\_tushar@sutd.edu.sg.}
\thanks{The authors are with the Singapore University of Technology and Design (SUTD), 8 Somapah Road, Singapore 487372. (Email: \{wayes\_tushar, lan\_lan, yunjian\_xu, chathura, yuenchau, kristinwood\}@sutd.edu.sg;  wang0555@e.ntu.edu.sg).}
}
\IEEEoverridecommandlockouts
\maketitle
\doublespace
\begin{abstract}
Air conditioning (AC) systems are responsible for the major percentage of energy consumption in buildings. Shared spaces constitute considerable office space area, in which most office employees perform their meetings and daily tasks, and therefore the ACs in these  areas have significant impact on the energy usage of the entire office building. The cost of this energy consumption, however, is not paid by the shared space users, and the AC's temperature set-point is not determined based on the users' preferences. This latter factor is compounded by the fact that different people may have different choices of temperature set-points and sensitivities to change of temperature. Therefore, it is a challenging task to design an office policy to decide on a particular set-point based on such a diverse preference set. As a consequence, users are not aware of the energy consumption in shared spaces, which may potentially increase the energy wastage and related cost of office buildings. In this context, this paper proposes an energy policy for an office shared space by exploiting an established temperature control mechanism. In particular, we choose meeting rooms in an office building as the test case and design a policy according to which each user of the room can give a preference on the temperature set-point and is ``paid'' for felt discomfort if the set-point is not fixed according to the given preference. On the other hand, users who enjoy the thermal comfort compensate the other users of the room. Thus, the policy enables the users to be cognizant and responsible for the ``payment'' on the energy consumption of the office space they are sharing, and at the same time ensures that the users are satisfied either via thermal comfort or through incentives. The policy is also shown to be beneficial for building management. Through experiment based case studies, we show the effectiveness of the proposed policy.
\end{abstract}
 \setcounter{page}{1}
\section{Introduction}\label{sec:introduction}
Air-conditioning (AC) systems are one of the primary sources of energy consumption,  and have a great influence on the overall energy usage in buildings. For example, it is estimated in \cite{BMS-Survey:2016} that buildings account for $70\%$ electricity consumption in the United States and $40\%$ of this consumption is due to the use of ACs~\cite{Wu_TCST_July2015}. Such massive consumption of energy can potentially increase the peak demand to the grid as well as significantly raise the cost of electricity for the building owner. As a result, there has been a considerable push towards designing policies and control algorithms that can effectively capture the tradeoff between an AC's temperature set-point and thermal comfort of users, and then control the ACs in such a way that confirms minimum inconvenience to the users~\cite{NHassan_TSG_Nov2015}.

The recent studies on policy design and control algorithms for controlling ACs can be divided into three categories. The first category includes policies for AC as demonstrated in \cite{Rocha_Feb2015,Hitchin_Jan2015} and \cite{Dyson_Oct2014}. In \cite{Rocha_Feb2015}, the authors present an intelligent optimisation model for deciding policies on energy sourcing via a dynamic temperature set point. A study to identify the potential impact of measures and policies to reduce the energy consumption of AC in European countries over a 10-year period is conducted in \cite{Hitchin_Jan2015}. Based on analysis of hourly electricity consumption data (smart meter data) from $30,000$ customer accounts in Northern California, suggestions are made in \cite{Dyson_Oct2014} to focus on ACs' energy consumption for designing policies and programs for demand response. A second category of studies mainly focuses on control strategies to minimize the use and cost of electricity. For instance, in \cite{Lee_TSG_Sep2013}, the authors propose an automatic thermostat control system based on the mobility prediction of users in indoor environments by using contextual information (historical pattern and route classification) through mobile phones. A combination of stochastic dynamic programming and rollout techniques is developed in \cite{Sun_TASE_July2013} for controlling ACs and lighting of buildings in order to minimize the total daily energy cost. The authors demonstrate a novel methodology of determining the proper chilled piping pressure set point in\cite{Su_TIA_June2013}, through which ACs used in high-tech industries can reduce electricity consumption. In \cite{Yu_TSG_Dec2013}, a mixed integer multi-scale stochastic optimization problem is formulated, and a model predictive control based heuristic is proposed for scheduling load of different characteristics including ACs for a home management system.

Finally, a third category of studies focuses not only on controlling the energy consumption for reducing cost but also on user comfort\cite{NHassan_TSG_Nov2015} and fault detection of ACs~\cite{Sun_TASE_Jan2014}. For example, a statistical process control technique and a Kalman filtering method are integrated in \cite{Sun_TASE_Jan2014} for system level fault detection in AC systems. A neural network based fault detection method is proposed in \cite{Du_EBE_Mar2014} that improves the energy efficiency and thermal comfort by removing various faults in the system. Two other robust and computationally efficient algorithms for fault detection techniques for AC are demonstrated in \cite{Bonvini_EAE_July2014} and \cite{Gao_EAE_2015}. These algorithms are based on a non-linear state estimation technique and a t-statistic approach respectively. Finally, user comfort has received paramount importance in a recent energy control study of AC systems. This attention is due to the fact if the user comfort is compromised significantly as part of an energy management system, consumers will eventually refuse to adopt such an management scheme~\cite{Tushar-TSG:2014}. In this regard, energy management techniques for ACs considering the user's comfort have been developed in \cite{NHassan_TSG_Nov2015,Zhou_TPS_May2015,Brundage_TASE_July2014,Javed_ITE_July2015,Mathews:2001,Canbay:2004,Anderson:2007,Brooks:2015,Chen:2014,Dobbd:2014,Atam:2015} and \cite{Lin_TSG_Mar2015}.

As shown from the above discussion, existing studies mainly focus on controlling ACs at private spaces (e.g., residential homes). However, ACs in shared spaces, e.g., in office space areas in which most office employees perform meetings and related activities, have a significant impact on the energy usage of a commercial building. Control of AC systems in shared spaces and the design of associate policies have not received significant attention.  This is mainly due to the diversity of users' preferences on the temperature set-point that are difficult to handle in deciding a set-point temperature, and  the lack of a pricing mechanism that can engage users for the payment of electricity they use. In this context, there is a need to develop solutions that can capture the conflicting interests or preferences of different users on the choice of temperature set-point in shared spaces, and can effectively benefit all the users in a fair fashion either in terms of thermal comfort or monetary revenue (either real or virtual) without increasing any cost to the building manager. To this end, this paper proposes a policy that can be implemented in a shared office space, e.g., meeting rooms, and can successfully address the above mentioned concerns.

While AC systems combined with user preferences have not been a focus of recent studies, we note that the control of appliances in shared spaces is not new, and extensive work has been done to investigate effective control of appliances in office space. Examples of such studies can be found in \cite{Agogino-AAAI:2004,Agogino-sensor:2006,Agogino-sensor:2008} and \cite{Agogino_Aug2011}. However, the basic difference between our work and these studies is that we focus on the control of an AC's set-point temperature, while these studies discuss the control of lights in shared spaces. Please note that there exists a number of lights in an office space. Hence, the change of light intensity of a space is decided by choosing different lights and optimally controlling their intensity as well as power consumption levels. Nonetheless, this paper focuses on AC in meeting rooms where it is considered that each meeting room has one AC. Therefore, control of the set-point temperature is more difficult as there is no other AC in the room to satisfy the preferences of occupants in a different zone of the room. Furthermore, the control of the AC associates a payment between different occupant, as we will see in the next section, which is another novel aspect of this designed policy.

To this end, the energy policy discussed in the paper considers an office building as the test case, where the design of the policy considers, on the one hand, each user of a room can provide a preference on the temperature set-point and is paid for felt discomforts if the set-point is not fixed according to the given preference. On the other hand, users who enjoy the thermal comfort compensate the other users of the room. Thus, the policy enables the users to be cognizant and responsible for the payment of the energy consumption of the office space they are sharing, and at the same time, ensures that the users are satisfied either via thermal comfort or through monetary incentives\footnote{A potential example of such monetary incentive could be in terms of virtual company money such as in \cite{PrintQuota:2013}.} (either real or virtual). The policy is also shown to be beneficial for building management. We understand that predicted mean vote (PMV) is another aspect that can be exploited to design a policy. Essentially, the PMV model relies on several comfort related environmental and personal parameters, such as air temperature, humidity, air speed, the clothing, and activity levels of occupants, among which \emph{air temperature} is mostly easily acquired and controlled in practical situations. Therefore, we concentrated our study of the policy on temperature. Nonetheless, we believe that with appropriate modification, the mechanism-based policy can further be generalized to incorporate some of the other parameters.

We run extensive experiments in a shared space environment of a tertiary education institute with $30$ participants who could report their preferences on the temperature set-point in real-time and then obtain the net-benefit in terms of thermal comfort and payment. The participants were students, researchers, and design engineers over $21$ years old recruited from an academic institution, including professionals from the staff and research centers. In each group there was a diversity in gender, age, and nationality\footnote{Please note that since the individual features of the participants were not the focus of the presented study, such features were not intentionally controlled during recruitment. Instead, a random sample of participants were chosen across gender, age, and nationality.}. In particular, we choose a meeting room to perform the experiment. Note that a building may consist of different types of shared spaces including meeting/conference rooms, business incubators, team space, work lounges, mail areas, pantry areas, break areas, and waiting areas, and any of these spaces has the potential to implement the proposed policy and thus contribute to energy savings for the buildings. However, meeting rooms are one of the most highly used areas within an office, and consume a substantial portion of total electricity of the entire building. In fact, a number of studies in the literature, such as \cite{MeetingScheduling:2016} and \cite{MeetingScheduling:2012}, has focused on different techniques for reducing electricity consumption in meeting rooms. As such, meeting rooms are a reasonable choice for showing the effectiveness of the proposed policy in this study. Through  case studies based on the data from the experiment, we show the effectiveness of the proposed policy. We note that the authors in \cite{Erickson:2012} also demonstrate an AC control mechanism in a shared space that considers the preferences of the occupants. However, although the baseline evaluation by taking votes from the participants follows the same procedure in \cite{Erickson:2012} and the proposed work, our policy substantially differs from \cite{Erickson:2012} in other aspects as discussed in Table~\ref{table:compare}.
\begin{table}[h]
\centering
\caption{Demonstration of the differences between \cite{Erickson:2012} and the proposed policy.}
\includegraphics[width=\columnwidth]{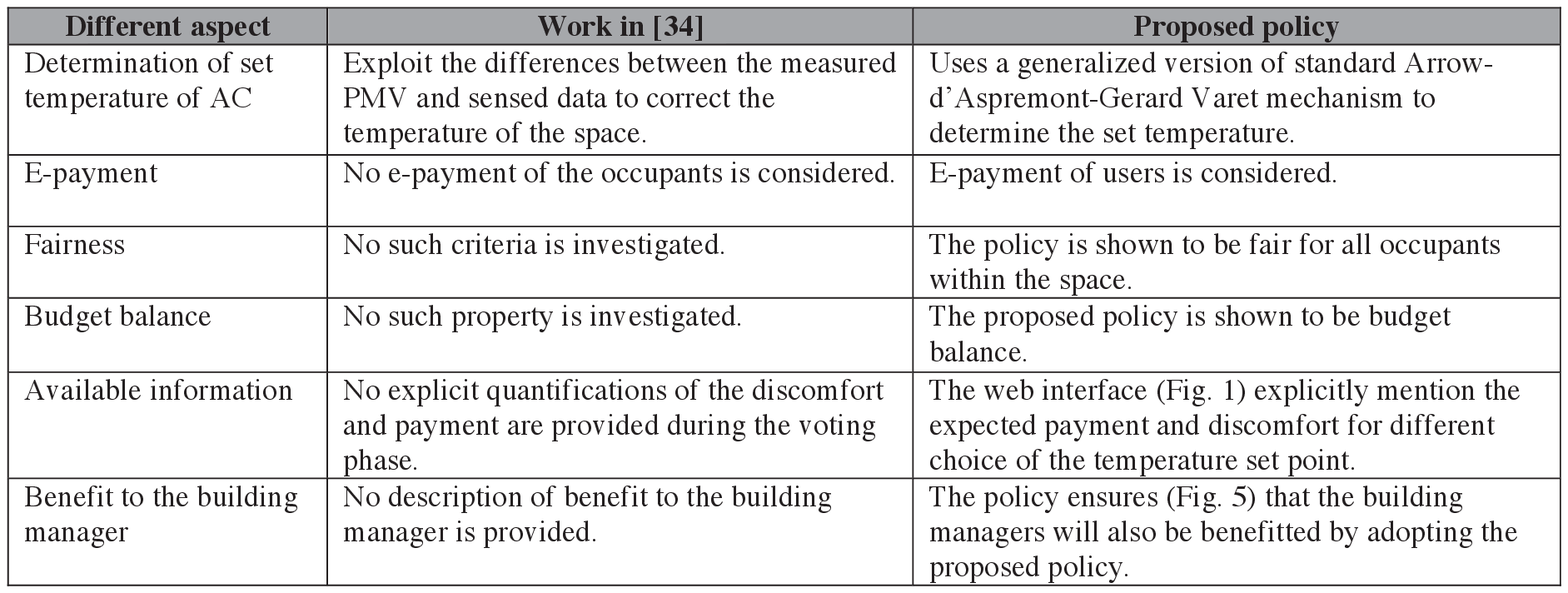}
\label{table:compare}
\end{table}
  
The remainder of the paper is organized as follows. A discussion on the mechanism that provides the foundation for the implementation of the proposed policy is discussed in Section~\ref{sec:section2}. Policies for AC temperature control in shared spaces are given in Section~\ref{sec:section3}. We provide exemplar physical and numerical case studies to show the beneficial properties of the policy in Section~\ref{sec:section4}. Finally, concluding remarks are drawn in Section~\ref{sec:section5}.

\section{Mechanism Behind the Policy}\label{sec:section2}
Due to the fact that the efficient use of ACs can significantly reduce the cost of energy, there have been a number of energy efficient AC systems\footnote{Please see the details on such AC systems in http://www.panasonic.com/sg/consumer/air-conditioners-learn/features-explanation/econavi-with-intelligent-eco-sensors.html} available in the market recently.  Nonetheless, in addition to the recent new AC technologies that can directly reduce an AC's power density, the consumption of electricity can further be reduced by effectively choosing the set-point temperature of the ACs. In this context, different incentive schemes have been developed to encourage the users to intelligently choose a set-point that can lead to potential energy savings in private space~\cite{NHassan_TSG_Nov2015}. However, because of diverse preferences and requirements on the room temperature by different individuals it is difficult to fix a temperature set-point in shared spaces.

To this end, we develop a mechanism (\cite{WangTao-CDC:2015}) by generalizing the standard Arrow-d'Aspremont-Gerard-Varet (AGV) mechanism. The objective is to develop an analytic solution that can chose the AC's temperature set-point in a shared space that finds the best trade-off between users' thermal comfort and energy cost. The benefit to an occupant is defined as the difference between the thermal comfort (or, discomfort) that the occupant feels from the set-point temperature of the room and the payment that she needs to pay (or receive) to keep that temperature. In other words, each occupant in the room is fairly benefited in terms of either thermal comfort or the payment, i.e., incentives, received due to feeling discomfort. To be more specific, the net benefit \(\pi_i(\bm\theta,\bm\alpha,\bm\beta)\) obtained by an occupant~\(i\) is defined as his/her valuation (for the selected outcome \(x\)) \(u_i(\theta_i,x)\) minus his/her payment \(t_i(\bm\theta,\bm\alpha,\bm\beta)\), i.e., \(\pi_i(\bm\theta,\bm\alpha,\bm\beta) = u_i(\theta_i,x) - t_i(\bm\theta,\bm\alpha,\bm\beta)\), where the payment is a function of all occupants' joint type \(\bm\theta\) as well as parameters \(\bm\alpha$ and $\bm\beta\).
In particular, the total benefit to all user is defined as a social welfare function
\begin{eqnarray}
W = \sum_{i=1}^n u_i \left(\theta_i,x \right)-C\left(x \right),
\label{eqn:equation1}
\end{eqnarray}
where the first term $\sum_{i=1}^n u_i\left(\theta_i, x \right)$ denotes the sum benefits to all the occupants of the room, the second term $C\left(x \right)$ refers to the cost associated with the chosen outcomes $x$ of the shared space occupant\footnote{Detail of $x$ is explained in Section~\ref{sec:section4}.}, and $u_i(\theta_i,x)$ is occupant $i$'s value (willingness to pay) for outcome $x$. $\mathbb{E}_\theta \left[\cdot\right]$ represents the expectation of $\left[\cdot\right]$ over the distribution of temperature $\theta$. Details on the welfare function and the associated distribution of social welfare can be found in \cite{WangTao-CDC:2015}.

It is important to note that the generalized AGV mechanism designed in   \cite{WangTao-CDC:2015} is shown to have several properties including incentive compatibility, efficiency, fairness and budget balance.
\begin{definition}
A mechanism is called incentive-compatible if every participant following the mechanism can achieve the best outcome to him/herself just by acting in accordance with his/her true preferences.
\label{definition:1}
\end{definition}
\begin{definition}
A mechanism is called efficient if the mechanism ensures that the sum benefits in terms of social welfare to all the participants following the mechanism is maximized.
\label{definition:2}
\end{definition}
\begin{definition}
Fairness refers to the property in which all participants following the mechanism are treated equally in terms of receiving the outcomes or benefits.
\label{definition:3}
\end{definition}
\begin{definition}
A budget balanced mechanism refers to a mechanism in which neither a budget deficit nor a budget surplus exists after the execution of the mechanism.
\label{definition:4}
\end{definition}

\begin{remark}
More details on the ``fairness" issue discussed in \cite{WangTao-CDC:2015} is described as follows. The concept of fairness is embodied by the relation among all occupants' net benefit \(\pi_i(\bm\theta,\bm\alpha,\bm\beta)\). Through manipulating the parameters $\bm\alpha$ and $\bm\beta$ two stages of ``fairness" can be achieved. At the first stage: the ex-ante net benefits of all occupants ${\mathbb E}_{\bm\theta}\{\pi_i\}$ are made to be equal to each other (${\mathbb E}_{\bm\theta}$ represents the expectation over the prior distribution of all occupants' types). However, although the ex-ante net benefits can be equal, the ex-post net benefits $\pi_i$ may be highly imbalanced. Therefore, at the second stage the parameters $\bm\alpha$ and $\bm\beta$ are further optimized in order to minimize the sum variance of all occupants' ex-post net benefits: \(\mathop {\min }\limits_{{\bm{\alpha }},{\bm{\beta }}} \sum\nolimits_{i = 1}^n {{{\mathbb E}_{\bm{\theta }}}{{\left( {{\pi _i}({\bm{\theta }},{\bm{\alpha }},{\bm{\beta }}) - {{\mathbb E}_\theta }\{ {\pi _i}({\bm{\theta }},{\bm{\alpha }},{\bm{\beta }})\} } \right)}^2}} \). Consequently, the ex-post net benefits are further balanced among all occupants, which reflects a concept of fairness.
\end{remark} 

As such, the designed mechanism in \cite{WangTao-CDC:2015} ensures that
\begin{enumerate}
\item All users within a shared space disclose their true preferences on the set-point temperature based on the payment (or, incentive) that they may need to pay (or, receive), which is a direct consequence of the \emph{incentive-compatible} property of the mechanism.
\item Since the mechanism is \emph{efficient}, the total benefit to all the room occupants in terms of social welfare (as shown in \eqref{eqn:equation1}) will be maximized if the mechanism is implemented for AC set-point temperature control in shared space.
\item Due to fairness, all the occupants of a shared space will be treated equally, and hence no occupant would feel inferior in terms of receiving the benefits from following a policy based on this mechanism.
\item The budget of the building manager will not be affected for setting different temperature set-points of a room's AC at different times based on the diverse preferences of the occupants by following the proposed mechanism in \cite{WangTao-CDC:2015} as the mechanism is \emph{budget balanced}.
\end{enumerate}

In this context, we propose a policy for controlling the temperature set-point of ACs in shared office spaces in the next section.
\section{Policy Design for AC Temperature Control}\label{sec:section3}
In this section, we outline the policies to control the temperature set-point of AC in a shared space. Please note that the policies are proposed based on the current AC control mechanism in office spaces, where tests have been performed in Singapore. In this discussion, first we enumerate the policies that the occupants and the owner of a shared space\footnote{For the particular policy, we assume meeting rooms as the shared space.} can adopt followed by a discussion on the rationale behind choosing the particular policy.
\subsection{Policies for AC control in Shared Space}
\begin{enumerate}
\item Anyone sharing a space will submit his/her choice on the temperature set point as soon as he/she enters the room (or provided equivalently from established preference profiles, and can change it at any point of time during his/her stay. The temperature set point, however, will be modified at every particular time interval set by the energy management system (EMS) with or without any manual control.
\item After each modification of temperature, an e-currency account of each occupant will either be debited or credited with virtual company money by an EMS based on his/her preferred temperature and the new temperature set point of the room.
\item If anyone forgets or decides not to provide any decision on his/her preference of the temperature set point, the EMS will create a preference on his/her behalf based on his/her historical preferences or preference profiles available in the system.
\item The EMS will pay for the energy consumed by the air conditioning and mechanical ventilation (ACMV) for the base temperature, which is chosen by the EMS based on regulations of the region.
\end{enumerate}

\subsection{Rationale Behind The Policy}\label{sec:rationale_policy}
As mentioned previously, the policies are designed to control the AC systems of shared spaces, where testing was performed in Singapore\footnote{The policies might need edits or modifications to apply in other countries.} with a view to benefit both the shared space owner and the occupants. To this end, the main rationale behind choosing the above mentioned policies are described as follows.
\begin{itemize}
\item This policy is designed to solve two problems that are particularly related to the use of ACMV in a shared space: deciding the temperature set point that takes into account the thermal comfort of all users using the shared space and a fair distribution of energy cost among the users. Therefore, this policy aims to maximize all users' net benefits, where each user's net benefit is considered as the difference between the user's valuation for the change in temperature and the received payment.
\item Every user is treated equally, i.e., no user has higher priority than any other user in the same space. Therefore, the users who decide and enjoy a specific temperature setting should compensate for others who suffer from a (uncomfortable) temperature in terms of monetary (incentive, real or virtual) transfer.
\item Virtual company money for changing users behavior towards using a shared facility has been extensively practiced in commercial buildings, e.g., see \cite{PrintQuota:2013} and \cite{PrintQuota:2014}. Therefore, payment of monetary compensation in terms of virtual company money can easily be leveraged in developing the proposed policy in shared space.
\item The difference of energy consumption caused by deviating from the base temperature\footnote{The cost of base temperature is covered by the company or building manager.} set point is fairly covered by all users according to their thermal preferences.
\item The decision making process of the EMS on the temperature set point and associated payments are designed using a generalized Aspremont-Gerard-Varet (AGV) as explained in Section~\ref{sec:section2}. Thus, due to the properties of the mechanism behind the decision making, the policies ensure that
\begin{itemize}
\item All users truthfully declare their choices on the preferred change of temperature set point.
\item The payment made by all users covers the difference of energy consumption cost due to their different preferences compared to a base temperature.
\item The set point of the ACMV is adjusted to a temperature at which the net benefit of all users is maximum.
\item Each user's expected net benefit is equal to other users in the same shared space.
\end{itemize}
\end{itemize}
In this context, now we describe the experimental procedure adopted in this study to set a preferred temperature for all the users such that fairness is maintained in terms of net benefits.

\section{Experiment Design and Case Study}\label{sec:experimental-set-up}
\subsection{Experiment Design}
We run a series of experiments to validate the effectiveness of the proposed policy so as to successfully adapt in a shared space of a building. In particular, the experiments are necessary to understand 1) what are the usual choices of preferred temperature by a majority of occupants in a shared space; 2) how sensitive an occupant is towards the temperature of a space (i.e., how much he or she intends to pay from his/her virtual account for keeping his or her preferred temperature); and 3) how the allocation of net-benefits can be made fair among all the occupants. Please note that unless we can address the above mentioned concerns, the outcome of the policy could be detrimental for the occupants and subsequently be rejected to implement in buildings' shared spaces. In this context, now we describe how the experiment was designed followed the procedure of conducting experiment at a meeting room in a tertiary educational institute. Then, we provide some numerical case study based on the experimental data to show that the experimental results, which mimic the implementation of the proposed policy, are beneficial for both the occupants and building management and ensure a fair allocation of net-benefit among the occupants based on their reported preferences.

The participants were allowed to bring their laptops (or tablets) to the experiment, and they were engaged in their work tasks either individually or as groups. Each experimental session was conducted in two phases: 1) Preference collection and 2) Fair allocation of benefits. \emph{Preference collection} phase was used to collect the participants' choices on the AC's set point based on the current room temperature. Please note that different activities may affect the participants' sensation for the environment and thus influence their willingness to express their preferences. However, our purpose of the experiment was to imitate a natural shared space where a diversity of activities could be observed. For example, we gave nine types of thermal comfort preferences (please refer to Table \ref{table:types}) for the participants to select from, which to some extent captured the random nature of participants' different thermal preferences. Such collection of preferences was necessary for determining the assortment of occupants' choices on the AC's temperature set-point. The pattern of each occupant's preferences captured at the first phase was stored in the system, and would be used as an input parameter to temperature determine policy at the second phase. Such information, moreover, can be exploited to \emph{initialize} the temperature preference of any new occupant\footnote{Such an occupant refers to a visitor who may temporarily occupy the space and does not have their preference information known to the building management.} in a shared space that can further be updated based on his or her preferences on the temperature. Furthermore, if a user forgets or decides not to provide his/her preference, the management system will be able to generate a preference type for the user according to his/her prior distribution that is available in the system. In the \emph{Fair allocation of benefits} phase, the participants not only could give their preferences on the temperature but also may have access to the information on the payment that they need to pay (or, receive) for keeping (or, leaving) their preferred temperature set-point. This phase of the experiment was conducted to verify the proposed policy discussed in Section~\ref{sec:section3} as well as to evaluate the \emph{fairness of the proposed policy} in distributing the net-benefits among the occupants.

\begin{figure}[t]
\centering
\includegraphics[width=0.8\columnwidth]{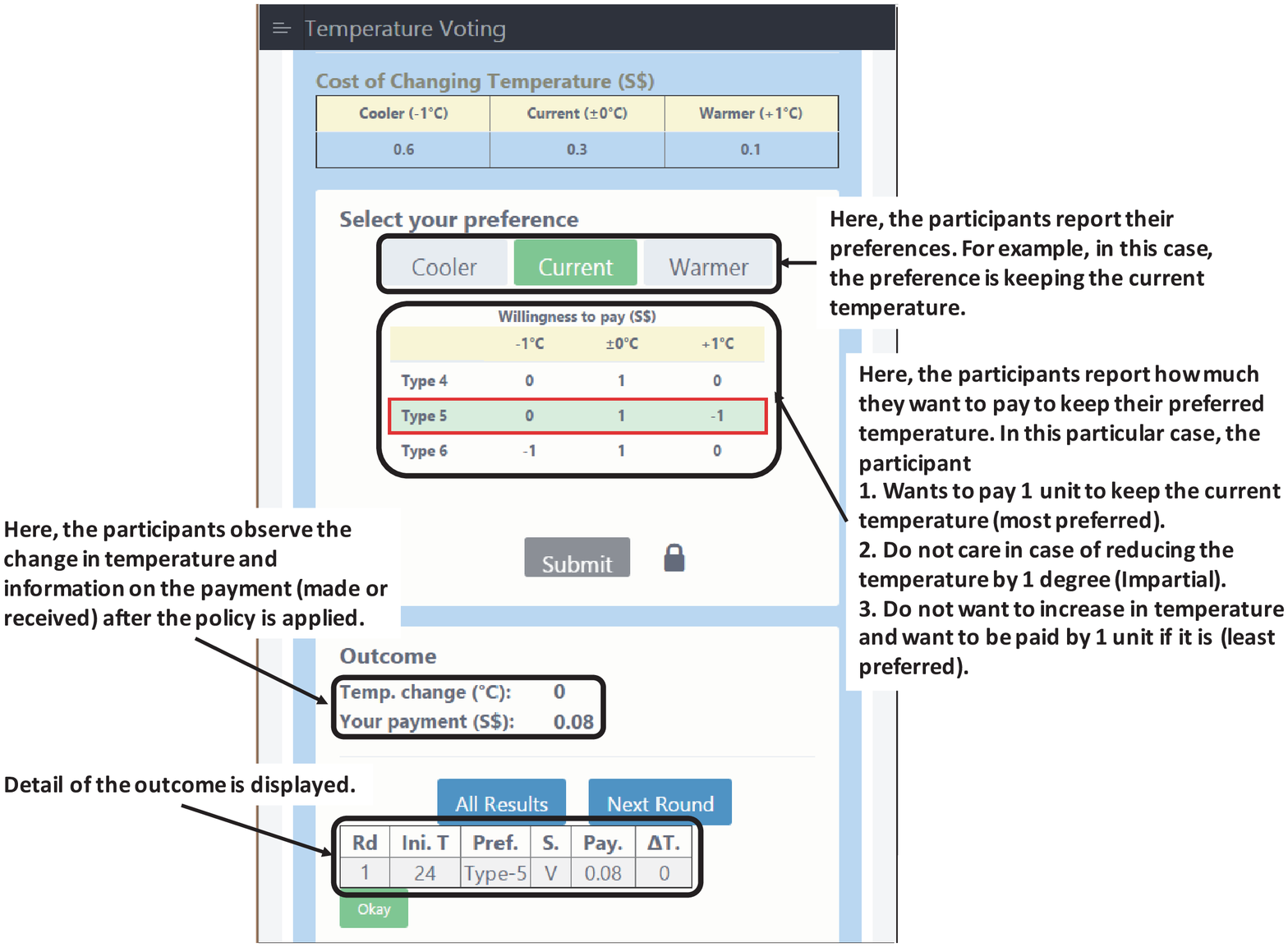}
\caption{Demonstration of the web-based interface that was used by each participant to report his/her preference on the temperature set-point of the AC. The monetary incentive (e.g., an equivalent electricity cost in this particular case) was for offering each participant a monetary value to motivate his/her decision making process on the temperature set point of the AC such that the participant felt appropriately being compensated for making his/her decision on the set point. The usefulness of such amount in influencing occupants' decisions was further verified through a survey.}
\label{fig:ReportingInterface}
\end{figure}

Each phase of the experiment was conducted for a duration of $2.5$ hours. Thus, the total duration of experiment was $5$ hours. During this $5$ hours, the participants' preferences were collected over $10$ reporting instances. Further, we ran the experiment within the temperature range of $[22^{\circ}{\rm C}, 26^{\circ}{\rm C}]$ with $1^{\circ}{\rm C}$ step size\footnote{In first phase, the temperature was monotonically increased from $22^{\circ}{\rm C}$ to $26^{\circ}{\rm C}$ with $1^{\circ}{\rm C}$ step size in every $30$ minutes, whereas in second phase the temperature was monotonically increased from $26^{\circ}{\rm C}$ to $22^{\circ}{\rm C}$ with $-1^{\circ}{\rm C}$ step size in same sampling rate.} (i.e., $5$ temperature points in total). Therefore, during the $5$ hour preference collecting phase, we were able to obtain two set of reporting data for each temperature point, and thus estimated initial prior distribution (over the $9$ possible preference types) for each user at each temperature point. The estimated prior distribution was subsequently used and further updated during the phase of \emph{fair allocation of benefits} of the experiment. Since it took around $15$ minutes for the meeting room temperature to reach a new set point, which was either $1^{\circ}{\rm C}$ more or less than the previous temperature, the occupants were asked to report their preferred temperature\footnote{The range of temperature for this experiment was chosen between $22^\circ\mathrm{C}$ and $26^\circ\mathrm{C}$.} of the room in every $30$ minutes using a web-based interface, as shown in Fig.~\ref{fig:ReportingInterface}, based on the current room temperature $T_0$. Here, the first $15$ minutes was used for the room temperature to reach the new set point, whereas in the next $15$ minutes the participants were allowed to feel and experience the new set temperature before asking them to submit their new sensation preferences. Please note that although the interface in Fig.~\ref{fig:ReportingInterface} displays a large quantity of information for the experimental purpose, it can be simplified substantially for shared space occupants during practical deployment. The thermal preferences were collected over three possible outcomes: i) reducing $T_0$ by one degree ($T_0 -1$), ii) maintaining $T_0$, and iii) increasing $T_0$ by one degree ($T_0 + 1$). Particularly for this experiment, each user reported one preference, i.e., how eagerly each participant wants his/her preference on set-point temperature to be implemented, out of a set of $9$ possible preferences as listed in Table~\ref{table:types}. It is important to note that the decision on the temperature set point of the AC, however, is determined based on the preferences made by all occupants within the room rather than depending on any individual preference. For example, if one of the occupants wants a warmer temperature, the policy does not necessarily set the temperature to a warmer value. This is due to the fact that the decision on the set point temperature is also affected by the other users in the group who, for example, agreed on another temperature rather than getting warmer.

We note that the requirement of occupants to log into a web interface upon entering a meeting space could be burdensome, and may frustrate their participation in submitting the comfort preferences. However, the manual input of preferences into the web interface has been used for the experiment only\footnote{\ds{For the experiment, we required the participants to report their preferences as a action throughout the experiment.}}, which can easily be extended to an automatic system of preference submission through the occupants' smart phone devices. For instance, each occupant can store his preferences within the server of the building management system (accessed via internet), which would be automatically read by the control server, and then the system would execute the decisions based on the designed policy as soon as it identifies the smart phone of the respective user within the meeting room.

Note that such an automatic system can easily be developed using existing technologies available in the market such as iBeacon \cite{iBeacon:2014}, which is a protocol developed by Apple through which portable devices (such as smart phones) can broadcast their identifier to nearby devices (e.g., EMS server in this case). Then, upon receiving the information on the occupant's identity within the meeting room, the control server can set the AC's set point and decide on the e-currency for the occupant following the proposed policy based on the predefined preferences of the smart phone holder. The preferences are stored within the control server, which can be accessed and changed by each respective occupant at any time via the Internet. Thus, such system allows the occupants to automatically participate in the policy without the need of manually providing the preferences to the system. Further, the building management system can also automatically get the identity of each occupant in the meeting room from their smart devices without needing them to input their identity in each time instant. Example different applications using iBeacon can be found in \cite{iBeacon:2016}.

\begin{table}[h]
 \centering\vspace{-2mm}
 \caption{This table demonstrates the types of thermal comfort preferences used in the experiment. Each row (except the last one) of column $1$ represents different type (i.e., $\theta_i$) of occupant, and the values on column $2$, $3$ and $4$ refer to willingness of the occupants to pay for an outcome $x$ (i.e., $u_i(\theta_i, x)$). The energy cost is demonstrated in last row of the table.}
    \begin{tabular}{ | c | c | c | c | c |}
        \hline
        \rowcolor{yellow!50} {Type} & $T_0-1$ & $T_0$ & $T_0+1$ & {Unit} \\ \hline\hline
        \rowcolor{blue!40} {Prefer cooler (1)} & $0.2$ & $0$ & $-0.2$ & {\$} \\ \hline
        \rowcolor{blue!40} {Prefer cooler (2)} & $0.4$ & $0$ & $-0.2$ & {\$} \\ \hline
        \rowcolor{blue!40} {Prefer cooler (3)} & $0.4$ & $-0.2$ & $-0.4$ & {\$} \\ \hline\hline
        \rowcolor{green!40} {Prefer current (1)} & $0$ & $0.4$ & $0$ & {\$} \\ \hline
        \rowcolor{green!40} {Prefer current (2)} & $0$ & $0.2$ & $-0.2$ & {\$} \\ \hline
        \rowcolor{green!40} {Prefer current (3)} & $-0.2$ & $0.2$ & $0$ & {\$}\\ \hline\hline
        \rowcolor{red!40} {Prefer warmer (1)} & $-0.2$ & $0$ & $0.2$ & {\$} \\ \hline
        \rowcolor{red!40} {Prefer warmer (2)} & $-0.2$ & $0$ & $0.4$ & {\$} \\ \hline
        \rowcolor{red!40} {Prefer warmer (3)} & $-0.4$ & $-0.2$ & $0.4$ & {\$} \\ \hline\hline
        \rowcolor{gray!40} {Energy Cost} & $C(1)$ & $C(2)$ & $C(3)$ & {\$} \\ \hline
    \end{tabular}\vspace{-1mm}
    \label{table:types}
\end{table}

In Table~\ref{table:types}, a positive value refers to the amount of money that a user would like to pay for keeping his/her preferred outcome (temperature setting). On the other hand, a negative value refers to the amount of money that the user would like to receive for not setting his/her preferred temperature as the set-point point temperature of the AC in the shared space. Further, the half-hour energy cost ($C(k),k=1,2,3$) incurred at each of the three outcomes is presented in the last row of Table~\ref{table:types}. In particular, the half-hour energy cost $C(k)$ (in \$) is estimated based on the energy consumption needed to maintain the required temperature set-point in the room. Note that the estimation of the energy consumption and subsequently the energy cost are necessary to determine the total budget that is required, in excess, to cover the cost of AC's electricity consumption resulting from different preference choices of the occupants. Furthermore, this budget is a critical input for the mechanism behind this policy as well as deciding the payment for the choice of each preference (detail can be found in \cite{WangTao-CDC:2015}).

\begin{figure}[t]
\centering
\includegraphics[width=0.7\columnwidth]{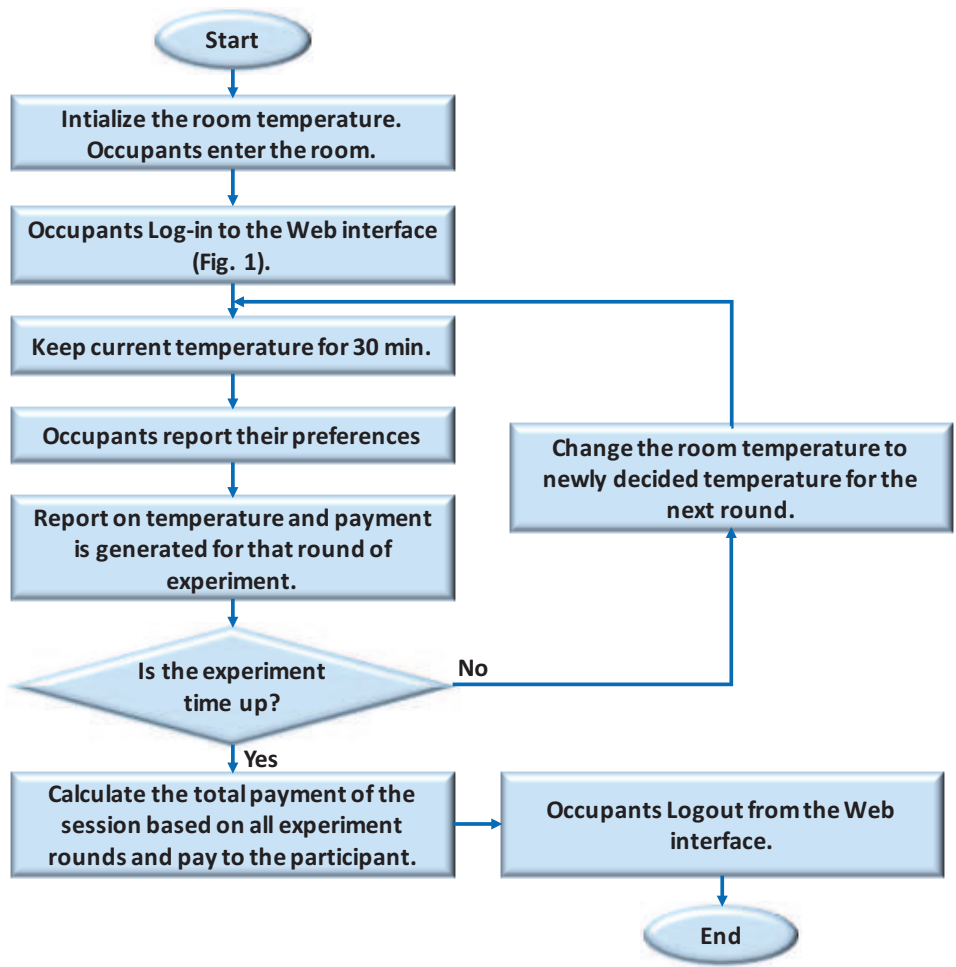}
\caption{Overview of the experiment procedure for the study of set-point temperature control of ACs at shared space.}
\label{fig:ExperimentProcess}
\end{figure}

Upon receiving the temperature preferences by the occupants and the electricity cost estimation by the energy model, the experiment coordinator who acts as the building management system used the generalized AGV mechanism~\cite{WangTao-CDC:2015} to determine the most suitable temperature of the room and the related payment to each participant such that the fairness between all the participants are guaranteed in terms of their received net benefits. Please note that, for experimental purposes, the mechanism behind the policy was run manually by the coordinator. Here, manual refers to the fact that each participant was asked by the coordinator of the experiment to give their preferences into the interface (Fig.~\ref{fig:ReportingInterface}), in intervals of every $30$ minutes. Then, after receiving all the preferences, the experiment coordinator gave them as input to the designed mechanism to decide on the new temperature set point of the meeting room and the monetary payment that was made to all the participants. However, the mechanism can easily be operated automatically if the policy is adopted in the shared space of a building as explained before. Thus, after the entire duration of the experiment (which can be referred to as the duration for which the occupants occupy a shared space) the total payment of each occupant was made in e-currency. Such e-currency could be in terms of virtual company money credited into or debited from the occupants' virtual account. For instance, each occupant's account can be credited with a certain amount of virtual company money for a particular duration of time (e.g. a month) that will enable the users to reserve and use different shared space within the buildings. The balance of the account will establish the priority of a user in reserving or using a space (higher balance refers to higher priority). Thus, with a higher priority users may enjoy in making reservation and use of different spaces within the building for different activities according to their preferences. However, if the account is empty or the balance is low, the user may have no decision making capacity on the choice of any space, and hence need to accept whichever space is offered to her by the system. As such, the users may be motivated to sincerely decide on the temperature setting of a shared space, if the policy is implemented, to keep the balance of their virtual account high. Please note the effectiveness of such debit and credit of virtual account on the users' behaviour towards efficient use of energy has also been demonstrated in \cite{PrintQuota:2014}. The overall experiment procedure is summarized in Fig.~\ref{fig:ExperimentProcess}.

\subsection{Case Study}\label{sec:section4}

\subsubsection{Experiment Set Up}
\begin{figure}[t]
\centering
\includegraphics[width=0.5\columnwidth]{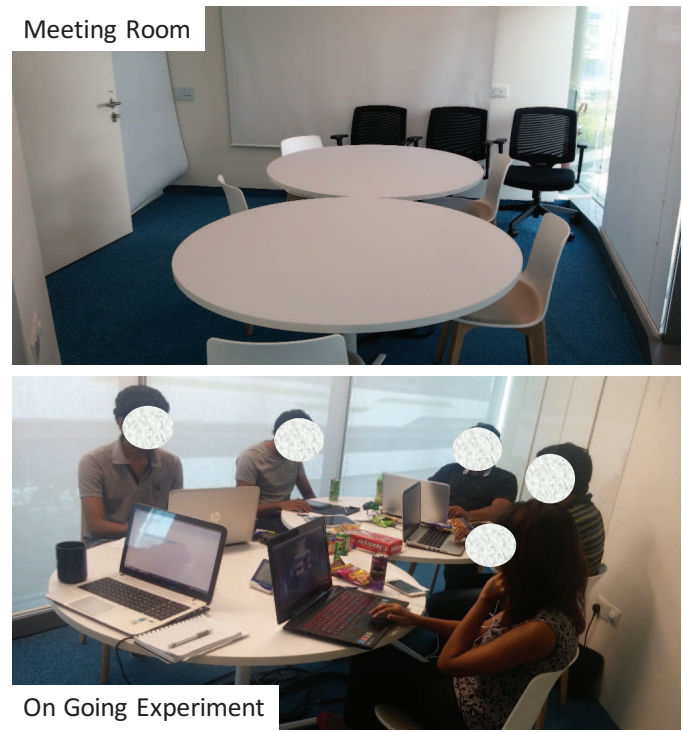}
\caption{Illustration of the meeting room that was chosen for running the experiment.}
\label{fig:MeetingRoom}
\end{figure}

The experiments were conducted in a meeting room of size $14.86$ m$^2$ that has a capacity of $6$ occupants. The room is located in a multi-story building in Singapore ($1.37^\circ$N, $103.98^\circ$E). The room is air-conditioned by a ACMV system that always operates in cooling mode due to the tropical climate of Singapore. The exterior surface is a fixed window with roller shades, and the shades were pulled down completely to cover the window during the entire experiment sessions. Besides the ACMV system, the major energy consumers in the meeting room include the lighting system (approximately 40 Watts) and the occupants' notebook computers (approximately 360 Watts). Please note that a meeting room was particularly chosen for the experiment due to the following reasons:
\begin{itemize}
\item The meeting room was equipped with necessary smart devices as the part of a smart energy project testbed that allowed us to collect and analyse the necessary data from the experiment.
\item The room was accessible for the entire duration of experiment, i.e., the meeting room was exclusively used for the experiment for one month, which we could not manage for other spaces within the building due to the conflict with schedules of other events of the facility.
\item Meeting rooms are one of the few shared spaces within the university which are equipped with fully standalone AC systems, which allowed us to change the set point temperature for this experiment at different times based on the proposed policy.
\end{itemize}
However, the same experimental set up can also be used for other shared space by altering some parameters such as the energy consumption model, as we will see shortly, of the set up. A picture of the meeting room used for the experiment, an on-going experiment and the detailed features of the considered room are shown in Fig.~\ref{fig:MeetingRoom} and Table~\ref{table:roomtype} respectively.

\begin{table}[h]
 \setlength{\arrayrulewidth}{0.5mm}
 \centering\vspace{-2mm}
 \caption{Architectural Features of the Considered Meeting Room}
 \rowcolors{2}{gray!40}{yellow!50}
    \begin{tabular}{ | l | l | }
        \hline
        \rowcolor{blue!40} {\bf{Building Component}} & {\bf{Meeting Room Model}} \\ \hline\hline
        {Location} & {Singapore}~($1.37^\circ\text{N},103.98^\circ\text{E}$)  \\ \hline
        {Floor} & {Single story}  \\ \hline
        {Floor Plan} & {Single room}  \\ \hline
        {Floor Area} & $14.56~\text{m}^2$  \\ \hline
        {Orientation} & {North}, $0^\circ$  \\ \hline
        {Lighting Power} & $40$~{Watts}  \\ \hline
        {Equipment Power} & $360$~{Watts}  \\ \hline
        {Occupant} & $6$~{Occupants}  \\ \hline
    \end{tabular}\vspace{-1mm}
    \label{table:roomtype}
\end{table}

\begin{figure}[t]
\centering
\includegraphics[width=0.7\columnwidth]{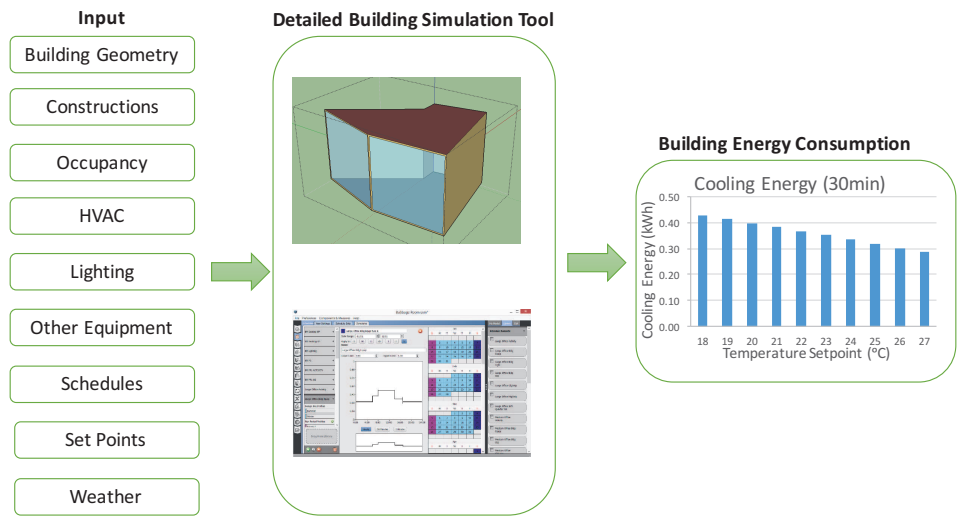}
\caption{Illustration of the energy model that has been used to estimate the half-hourly energy consumption of the meeting room during the experiment.}
\label{fig:EnergyModel}
\end{figure}

To participate in the experiment, $30$ participants were recruited, where we divided the participant pool into $6$ groups. Hence, each group of participants consisted of $5$ people, who along with an experiment coordinator (i.e., $6$ people in total) occupied the room during each experimental sessions. 
For this experiment, the energy consumption, which is used to calculate the half-hour energy costs $C(k)$ as aforementioned, is estimated using a building energy simulation program EnergyPlus. An ideal Variable Air Volume (VAV) terminal unit in the EnergyPlus module, i.e., ZoneHVAC:IdealLoadsAirSystem~\cite{EnergyPlus}, was used to describe the operation of the ACMV system. Then, with real-time weather data from an on-site weather station, determined every $30$ minutes during the experiment, we run the EnergyPlus simulation to calculate the half-hour cooling load for each of the three possible outcomes at the current temperature $T_0$. Finally, the obtained cooling load is translated into electricity consumption using a constant coefficient of performance (COP)~\cite{WJCole:2013}.
A demonstration of the model used for estimating energy consumption in the experiment is shown in Fig.~\ref{fig:EnergyModel}

\begin{figure}[h]
\centering
\includegraphics[width=0.5\columnwidth]{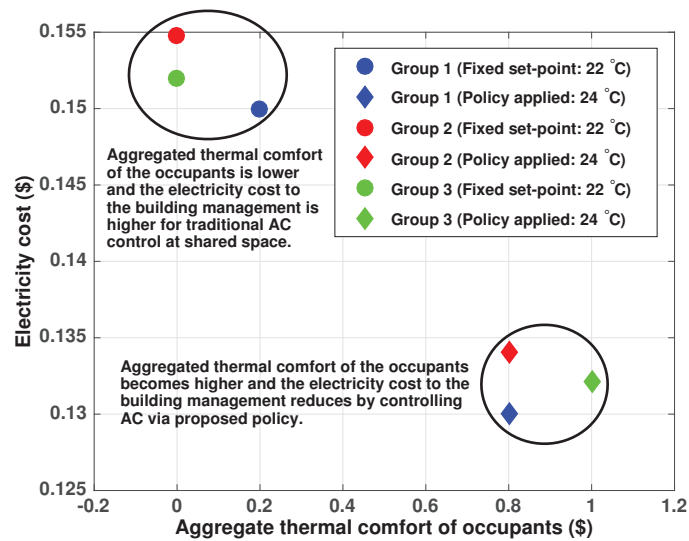}
\caption{This figure shows how the proposed policy works through choosing the outcome that maximizes the social welfare. It Illustrates the aggregate sum utilities of participants of the group $1, 2$ and $3$ and the related energy costs to the building management under fixed set-point scheme and the proposed policy. Similar results were also found for participants in group $4, 5$ and $6$.}
\label{fig:aggregate_utility}
\end{figure}

Here, it is important to note that the EnergyPlus model would need to be created once for a building to implement the policy. Once the model is implemented, most of the parameters related to the model remain unchanged. However, information on weather and occupants inside the building varies across time, which can live feed into the model directly using wireless sensor and an IoT (internet-of-things) network. Thus, once a model is created for a building, it can be used for all the time unless there is a significant change in the building infrastructure. Further note that the simulation platform for EnergyPlus is free online and is compatible with all available operating systems including Windows, Linux and Mac, and the implementation is not time consuming\footnote{See: http://apps1.eere.energy.gov/buildings/energyplus/.}. However, alternative method such as benchmarking against similar type of building (e.g., by using benchmarking report such as \cite{BenchMark_2014}) is also possible to adopt for determining the energy consumption of HVAC with different temperature setpoints. However, although such benchmarking technique would be less accurate than the EnergyPlus model. Another potential choice could be to use Chiller water flow data to estimate the energy consumption and related cost. Nonetheless, our motivation behind using the EnergyPlus model over Chiller water flow are provided as follows.

\begin{itemize}
\item Although chiller water flow data provides information on the real-time electricity consumption of a HVAC system, it does
not provide the estimation of energy consumption for different temperature set-points. Therefore, designing a model based on the chiller water flow data would require to run the HVAC system for a set of large number of different temperature set points and then use the data for developing the model, which could be troublesome and time consuming. Further, considering the fact that most commercial buildings have centralized HVAC system, it would be difficult to change the HVAC set point of the building centrally to different temperature to collect the data.
\item EnergyPlus model provides us with flexibility to predict the energy consumption and related cost of HVAC system that are not only limit to Chiller plant. Essentially, EnergyPlus model can be used for any HVAC system that are currently available in the market.
\end{itemize}

\subsubsection{Experiment Results}
In this section, we demonstrate the effectiveness and beneficial properties of the proposed policy. To do so, we conduct a numerical case study based on the preference data that was collected from the physical experiment described in the previous section. Considering the thermal comfort preferences, first we show how the designed policy can benefit both the shared space occupants and the building management in terms of aggregated thermal comfort and electricity cost respectively. To this end, we represent the aggregate thermal comfort of all occupants at \(T_0\) as \(\sum_{i=1}^n u_i(\theta_i,x)\), where $x$ denotes the chosen outcome from $T_0-1$, $T_0$, and $T_0+1$.  Here, all the valuation (willingness to pay) are drawn from a range with a maximum and minimum values of $0.4$ and $-0.4$ respectively (see Table \ref{table:types}). Hence, for each group of $5$ participants, the possible aggregated thermal comfort lies within the range of $\left[-2, 2\right]$. The relation between the aggregated thermal comfort of the users and the energy cost incurred to the building management is shown in Fig.~\ref{fig:aggregate_utility}. We illustrate this result in two subfigures in order to clearly show the benefits to all six groups that participated in the experiment. In demonstrating the results in Fig.~\ref{fig:aggregate_utility},
\begin{itemize}
\item The temperature of the room was initially set to $22^\circ\mathrm{C}$. (Due to the tropical climate of Singapore, a room temperature of $22^\circ\mathrm{C}$ sits at the lower side of the thermal comfort zone, and most participants will prefer higher temperature than this temperature. Therefore, starting at $22^\circ\mathrm{C}$ is a reasonable choice, which would potentially lead to a decision of a higher temperature and thus benefiting energy conservation).
\item The preferences of the participants were collected following Fig.~\ref{fig:ExperimentProcess}, and a new AC set-point temperature was established by the designed policy.
\item At the end of the experiment, the stable temperature was found to be $24^\circ\mathrm{C}$.
\end{itemize}
As shown in Fig.~\ref{fig:aggregate_utility}, for the set-point temperature of $24\,^{\circ}\mathrm{C}$, the aggregate comfort level attains a higher value while a lower electricity cost is incurred to the building management if the designed policy is implemented. For example, consider the case of group $1$ in Fig.~\ref{fig:aggregate_utility}. The aggregated thermal comfort\footnote{An aggregate thermal comfort of one unit with respect to a certain temperature setting refers to the total willingness to pay an amount of one $\$$ by a group of participants.} that the group achieved from implementing the policy increases from $\$ 0.2$ to $\$ 0.8$ ($4$ times increase) whereby the cost to the building management is reduced from $\$ 0.15$ to $\$ 0.13$  (which demonstrates a reduction of $13.33\%$ of cost). According to Fig.~\ref{fig:aggregate_utility}, the pattern is similar for other groups of participants as well.

Although the aggregated thermal benefits to the occupants in a shared space is maximized by adopting the policy, as explained in Section~\ref{sec:section3}, there is no guarantee that each occupant will experience his/her preferred thermal comfort as a result of this policy. This is due to the fact that a shared space accommodates different types of people with different thermal sensitivities. As a result, their preferences for set-point temperature of the room may deviate significantly from one another. Hence, the policy, as discussed in  Section~\ref{sec:section3}, compensates (credits) the occupants with relatively high thermal discomfort (comfort) with e-currency, and the net benefit of the policy is obtained as a difference between the thermal comfort and the payments (as shown in \eqref{eqn:equation1}). Nonetheless, it is significantly important that the net benefit that each shared space occupant achieves as a result of this policy needs to be fair and treats all of the occupants equally. Otherwise, the effectivity of the policy will be compromised, and eventually the policy will be rejected to be implemented in practical settings.

\begin{figure}
\centering
\includegraphics[width=0.6\columnwidth]{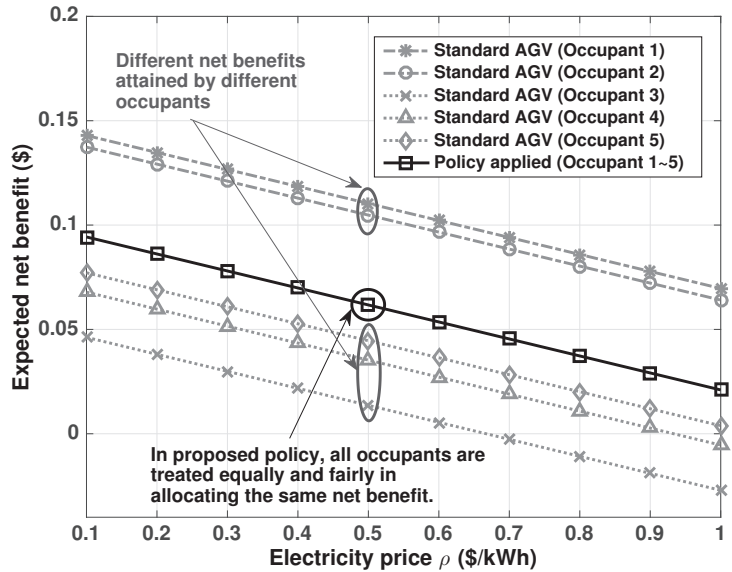}
\caption{The demonstration of the expected net benefits of participants in a group (here, the results are shown for group 2) under different electricity price $\rho$. The current temperature is set to be $24\,^{\circ}\mathrm{C}$. Here, net benefit refers to a user's valuation for the chosen outcome minus his/her payment as explained in Section~\ref{sec:rationale_policy}. As can be seen from the figure, the proposed policy treats all occupants equally, and fairly allocates the same net benefit to each of them.}
\label{fig:fairness}
\end{figure}

In this context, to illustrate the fairness property of the policy, we compare the proposed policy, which is based on the generalized AGV mechanism in \cite{WangTao-CDC:2015} with the policy based on standard AGV mechanism, in terms of the expected net benefits\footnote{Note that while aggregated thermal comfort (Fig.~\ref{fig:aggregate_utility}) represents the sum of all users' \emph{valuation} as a group of a specific outcome, net benefit is associated with a single user's \emph{valuation and payment}.} received by users of each group. The users' expected benefits of all participants in a group are shown in Fig.~\ref{fig:fairness}. In this figure, we vary the energy cost $C(k)$ by tuning the electricity price $\rho$. As shown in the figure, the same expected net benefit is attained by all participants (in a group) as a result of the proposed policy. Participants' expected benefits given by the standard AGV mechanism, however, usually deviate significantly from each other. For example, at a price of $0.5~\$$/kWh, the expected net-benefit of each occupant varies from each other significantly for the policy using the standard AGV mechanism. For instance, occupants $1$ and $2$ have higher benefits (more than $0.1$), whereas occupants $3, 4$ and $5$ have net benefits less than $\$0.06$. The proposed policy, however, treats all the occupants equally and allocates a net benefit of $\$0.062$ to each of them. Thus, the fairness of the net benefits allocation is established. Please note that only one plot is shown for the proposed policy in Fig.~\ref{fig:fairness}. This is due to the fact that all users received the same expected net benefit at different electricity price. As a result, the expected benefit of all occupants overlapped with one another.

Another important feature of Fig.~\ref{fig:fairness} is that the expected net benefit of each occupant decreases with the increase of electricity price. This is mainly due to the budget balance property of the policy where the expected net benefit is equal to the expected social welfare as defined in Eq. \eqref{eqn:equation1} (for details, please see \cite{WangTao-CDC:2015}). Therefore, it is obvious from Eq. \eqref{eqn:equation1} that with a higher electricity price, although the electricity cost $C(x)$ will increase, there will be a decrease in social welfare. As a consequence, users' expected net payoffs tend to decrease as well.

\begin{figure}
\centering
\includegraphics[width=0.6\columnwidth]{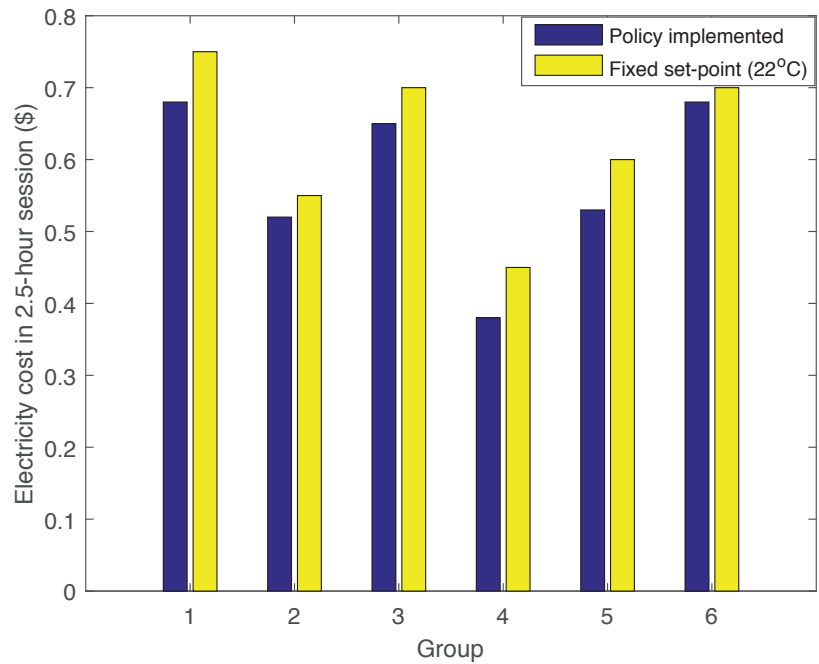}
\caption{Comparison of the cost of electricity consumed by the AC between the proposed and the scheme with fixed set-point temperature for each experimental session with the group.}
\label{lab:figure1}
\end{figure}
Finally, in Fig.~\ref{lab:figure1}, we compare the cost of electricity consumption by the AC in the meeting room for the proposed policy with the cost for the scheme with fixed temperature set point. According to this figure, with implementing the policy, the temperature is usually turned to be higher than the (conservatively low) fix-set point, and thus energy saving is achieved (while AC works in a cooling mode). The energy costs are estimated using the EnergyPlus model. However, the energy consumption for the fixed set-point scheme varies from group to group. This is due to the fact that the experiments with different groups were run at different times, and thus the consumption was largely affected by outdoor weather conditions. According to Fig.~\ref{lab:figure1}, compared to the fixed set-point scheme, the average energy saving across the groups is about $9\%$.

\section{Conclusion}\label{sec:section5}
In this paper, a policy is proposed to control the set-point temperature of AC systems in shared spaces. In contrast to existing AC control practices, the policy is designed to associate and establish the responsibility of the electricity usage of AC systems to the occupants of a space so as to make them aware of the cost of electricity wastage. The policy is based on an established temperature control mechanism in which each occupant can input their preferences on the set-point temperature. The net benefits in terms of experienced thermal comfort and associated cost is fairly allocated among the occupants. The effectiveness of the designed policy is established through experiments, and it is shown that the policy can benefit occupants by increasing the aggregated thermal comfort that they experience, while maintaining a fair distribution of net benefit among each occupant. Further, the policy is shown to be beneficial for the building management in terms of reducing the cost for maintaining a set-point temperature of AC at a preferred stable temperature decided by the occupants through the designed policy.

One potential extension of the proposed work could be the inclusion of various other parameters such as insulation level of the occupants' clothing or the pattern of their usage of different clothings within the policy to decide on the set point temperature of the shared space.

\section*{Acknowledgement}This work is supported in part by the Singapore University of Technology and Design (SUTD) under grant NRF2015ENC-GBICRD001-028 and NRF2012EWT-EIRP002-045, and in part by the SUTD-MIT International Design Centre (IDC;idc.sutd.edu.sg). Any findings, conclusions, recommendations, or opinions expressed in this document are those of the authors and do not necessary reflect the views of the IDC.

\end{document}